\documentclass[aps,prd,twocolumn,superscriptaddress,preprintnumbers,nofootinbib]{revtex4-1}

\usepackage{graphicx}
\usepackage{bm}
\usepackage{times}
\usepackage{slashed}
\usepackage{amssymb}
\usepackage{color}
\usepackage{slashed}
\usepackage{lipsum}
\usepackage{subfigure}
\usepackage{multirow}
\usepackage{amsmath}
\usepackage{array}
\usepackage{varwidth}
\usepackage{float}

\begin{document}

\title{Initial performance of the Radar Echo Telescope for Cosmic Rays, RET-CR}

\author{P.~Allison}
\affiliation{Department of Physics, Center for Cosmology and AstroParticle Physics (CCAPP), The Ohio State University, Columbus OH 43210, USA}
\author{J.~Beatty}
\affiliation{Department of Physics, Center for Cosmology and AstroParticle Physics (CCAPP), The Ohio State University, Columbus OH 43210, USA}
\author{D.~Besson}
\affiliation{University of Kansas, Lawrence, KS 66045, USA}
\author{A.~Connolly}
\affiliation{Department of Physics, Center for Cosmology and AstroParticle Physics (CCAPP), The Ohio State University, Columbus OH 43210, USA}
\author{A.~Cummings}
\affiliation{Departments of Physics and Astronomy \& Astrophysics, Institute for Gravitation and the Cosmos,   Pennsylvania State University, University Park, PA 16802, USA}
\author{C.~Deaconu}
\affiliation{Enrico Fermi Institute, Kavli Institute for Cosmological Physics, Department of Physics, University of Chicago, Chicago, IL 60637, USA}
\author{S.~De~Kockere}
\affiliation{Vrije  Universiteit  Brussel, Dienst ELEM, IIHE,  Brussels,  Belgium}
\author{K.D.~de~Vries}
\affiliation{Vrije  Universiteit  Brussel, Dienst ELEM, IIHE,  Brussels,  Belgium}
\author{D.~Frikken}
\affiliation{Department of Physics, Center for Cosmology and AstroParticle Physics (CCAPP), The Ohio State University, Columbus OH 43210, USA}
\author{C.~Hast}
\affiliation{SLAC National Accelerator Laboratory, Menlo Park, CA 94025, USA}
\author{E.~Huesca~Santiago}
\affiliation{Vrije  Universiteit  Brussel, Dienst ELEM, IIHE,  Brussels,  Belgium}
\author{C.-Y.~Kuo}
\affiliation{National Taiwan University, Taipei, Taiwan}
\author{A.~Kyriacou}
\affiliation{University of Kansas, Lawrence, KS 66045, USA}
\author{U.A.~Latif}
\affiliation{Vrije  Universiteit  Brussel, Dienst ELEM, IIHE,  Brussels,  Belgium}
\author{J.~Loonen}
\affiliation{Vrije  Universiteit  Brussel, Dienst ELEM, IIHE,  Brussels,  Belgium}
\author{I.~Loudon}
\affiliation{Universit\'{e} Libre de Bruxelles, Brussels, Belgium}
\affiliation{Department of Astrophysics/IMAPP, Radboud University, P.O. Box 9010, 6500 GL Nijmegen, The Netherlands}
\author{V.~Lukic}
\affiliation{Vrije  Universiteit  Brussel, Dienst ELEM, IIHE,  Brussels,  Belgium}
\author{C.~McLennan}
\affiliation{University of Kansas, Lawrence, KS 66045, USA}
\author{K.~Mulrey}
\affiliation{Department of Astrophysics/IMAPP, Radboud University, P.O. Box 9010, 6500 GL Nijmegen, The Netherlands}
\author{J.~Nam}
\affiliation{National Taiwan University, Taipei, Taiwan}
\author{K.~Nivedita}
\affiliation{Department of Astrophysics/IMAPP, Radboud University, P.O. Box 9010, 6500 GL Nijmegen, The Netherlands}
\author{A.~Nozdrina}
\affiliation{University of Kansas, Lawrence, KS 66045, USA}
\author{E.~Oberla}
\affiliation{Enrico Fermi Institute, Kavli Institute for Cosmological Physics, Department of Physics, University of Chicago, Chicago, IL 60637, USA}
\author{S. Prohira}
\affiliation{University of Kansas, Lawrence, KS 66045, USA}
\author{J.P.~Ralston}
\affiliation{University of Kansas, Lawrence, KS 66045, USA}
\author{M.F.H.~Seikh}
\affiliation{University of Kansas, Lawrence, KS 66045, USA}
\author{R.S.~Stanley}
\affiliation{Vrije  Universiteit  Brussel, Dienst ELEM, IIHE,  Brussels,  Belgium}
\author{S.~Toscano}
\affiliation{Universit\'{e} Libre de Bruxelles, Brussels, Belgium}
\author{D.~Van~den~Broeck}
\affiliation{Vrije  Universiteit  Brussel, Dienst ELEM, IIHE,  Brussels,  Belgium}
\affiliation{Vrije  Universiteit  Brussel, Astrophysical Institute,  Brussels,  Belgium}
\author{N.~van~Eijndhoven}
\affiliation{Vrije  Universiteit  Brussel, Dienst ELEM, IIHE,  Brussels,  Belgium}
\author{S.~Wissel}
\affiliation{Departments of Physics and Astronomy \& Astrophysics, Institute for Gravitation and the Cosmos,   Pennsylvania State University, University Park, PA 16802, USA}
\collaboration{Radar Echo Telescope}

\begin{abstract}
    The Radar Echo Telescope for Cosmic Rays (RET-CR), a pathfinder instrument for the radar echo method of ultrahigh energy (UHE) neutrino detection, was initially deployed near Summit Station, Greenland, in May 2023. After a 4 week commissioning period, 9 days of data were taken before the instrument went offline. In this article, we describe the instrument as it was deployed, and the initial performance of the detector. We show  that the technical aspects of running a radar based particle cascade detector in the ice have been demonstrated. Analysis of the 2023 data informed improvements that were incorporated into the May-August 2024 deployment, which has just concluded at time of writing. Results from the 2024 run will be presented in forthcoming publications. 
\end{abstract}

\maketitle
\section{Introduction}
The Radar Echo Telescope for Cosmic Rays (RET-CR)~\cite{retcr} is a pathfinder instrument for the radar echo method of ultrahigh energy (UHE) neutrino detection. When a UHE particle initiates a cascade within a dense material, a short lived cloud of ionization is created that can reflect incident radio waves. With the radar echo method, a transmitting antenna and receiving antenna(s) are deployed within a large target volume, such as the large ice sheets found in the Arctic and Antarctic, to detect the radar reflections from neutrino initiated cascades. For more information on the history, theory, and status of the method, see Refs.~\cite{chiba,krijnkaelthomas,radioscatter,t576_run2, retcr}. 

\begin{figure}[h]
  \centering
  \includegraphics[width=0.4\textwidth]{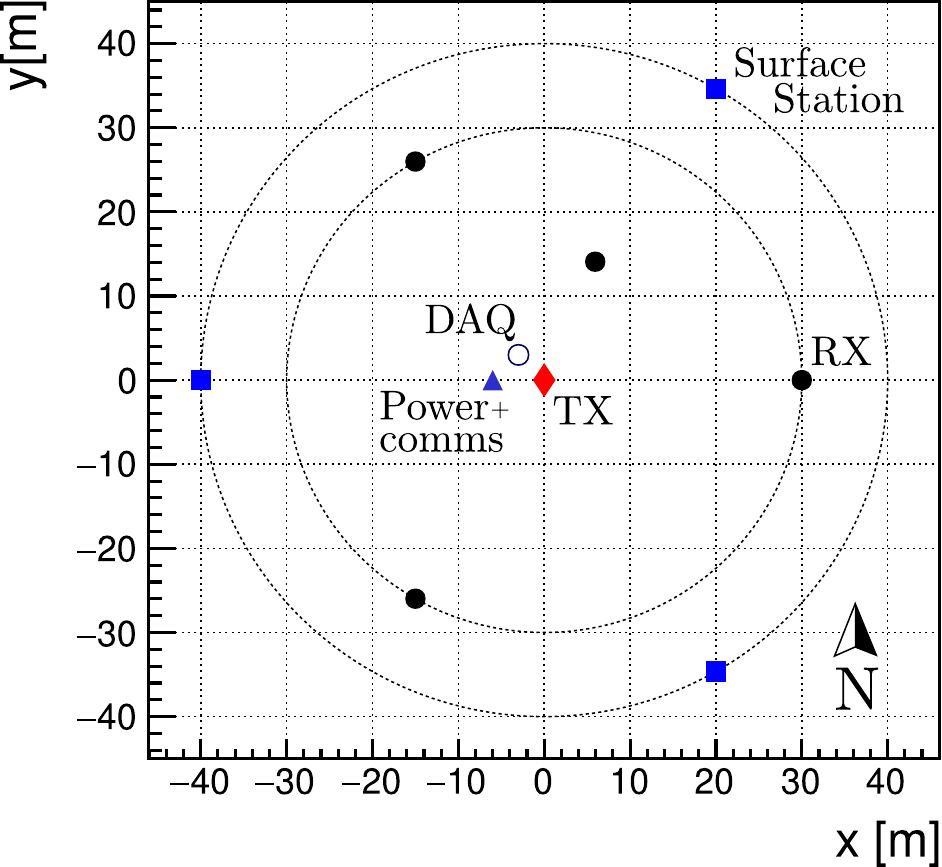}
    \caption{RET-CR layout as deployed in May 2023. Different system components (surface stations, data acquisition system [DAQ], transmitter [TX] and receiver [RX]) are indicated on the figure, and described in the main text.}
  \label{layout}
\end{figure}

RET-CR was developed to test the radar echo method in nature and demonstrate the technology needed to detect neutrino initiated cascades. RET-CR targets the in-ice component of a UHE cosmic ray extensive air shower (EAS)~\cite{eas}. At zenith angles less than $\sim$30 degrees and energies above $\sim$10\,PeV a significant fraction, $>$10\%, of the primary energy is deposited into the ground within a few cm of the EAS axis~\cite{DeKockere:2022bto}. This produces a secondary, dense cascade that mimics a neutrino initiated cascade, but at a far higher event rate. Targeting UHE cosmic rays is therefore a bridge between the successful laboratory detection~\cite{t576_run1,t576_run2} and an eventual neutrino detector~\cite{Ackermann:2022rqc}.

In this article we detail the detector as deployed at Summit Station, Greenland in May 2023, the deployment itself, and the initial performance of the instrument. We begin by discussing the differences between what was described in Ref.~\cite{retcr} and what was actually deployed; primarily a difference in layout geometry and number of stations. We then describe the deployment itself, as well as the commissioning phase data run. We then describe the performance of the various subsystems, and conclude with a discussion of the upgrades and improvements that were realized for the full data run in 2024.

\section{Description of the instrument} RET-CR can be divided into two main systems, the surface component and the radar component. The surface component consists of autonomous, solar powered stations, each with a pair of scintillator detectors and a dual polarized radio antenna. The radar component consists of a solar power system, a communications system, a central data acquisiton system (DAQ), a radar transmitter, and radar receivers. The principle of operation of the instrument is unchanged from what was presented in Ref.~\cite{retcr}, though some technical aspects are detailed below. The entire instrument was comprised of three surface stations, three downhole receiver strings, and one downhole phased transmitter string, in a layout shown in Fig.~\ref{layout}.

RET-CR operates via triggers from the surface system. A surface station will send a trigger to the radar system when it registers a coincident minimally ionizing particle (MIP) in both panels. The window for coincidence is 1\,$\mu$s. The radar system triggers an event readout when $\geq$N surface stations trigger within a 1\,$\mu$s window, with N being set in software. The radar system can also be triggered via software, GPS, or via a selectable edge threshold in the receiver (RX) channels.

\begin{figure}[h]
  \centering
  \includegraphics[width=0.5\textwidth]{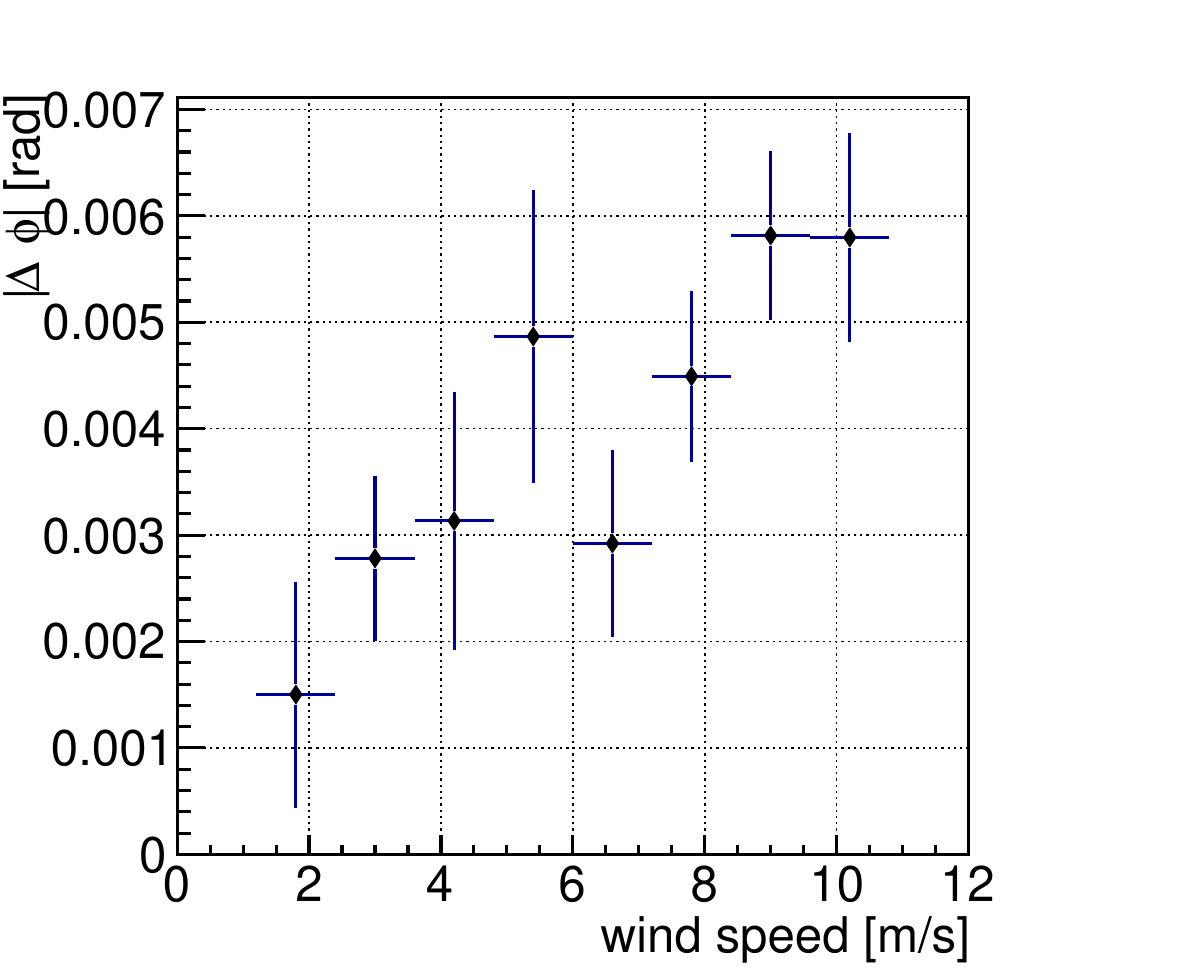}
    \caption{The magnitude of the change in phase required to cancel the TX signal correlates with wind speed.}
  \label{windspeed}
\end{figure}

\subsection{Radar System} The radar system is an 8 channel phased transmitter with 4 receive channels featuring active transmitter cancellation (TC). The heart of the DAQ is a xilinx RFSoC, featuring an 8 channel analog-to-digital converter (ADC) and an 8 channel digital-to-analog converter (DAC). The RFSoC was clocked at $\sim$~3GHz. A custom-made board breaks out all 8 DAC channels and all 8 ADC channels. The DAC channels have a software controlled variable attenuator providing up to 32\,dB of attenuation per channel. Four of the DAC channels are used to actively cancel the TX signal in the RX channels by injecting a scaled, time-delayed copy of the transmitted signal into the RX signal path. Overall attenuation of the transmitted signal in the RX of $>$80\,dB is achieved. The TX and RX antennas are all identical wide-cylinder dipoles with a 6.35\,cm radius. For the 8-channel TX phased array, each antenna is fed by a single 20\,W power amplifier, and each pair of power amplifiers is fed by a single channel of the DAC, allowing for pairwise phasing of the transmitted polar angle. The 8 TX antennas are separated by $\sim$0.3\,m\footnote{The wavelength of radio changes with depth owing to the changing density and index of refraction of ice, but we ignore these changes over the $\sim$5\,m length of our deployment string.}, deployed such that the midpoint between the fourth and fifth antennas is approximately 10\,m deep. Each RX channel consists of an antenna, a combiner for the transmitter cancellation, a $\sim$60\,dB low-noise amplifier, and a bandpass filter from 100-300\,MHz. Each RX is lowered down a borehole to a depth of 10\,m a distance of 30\,m from the transmitter (TX) borehole. LMR-400 coaxial cable was used for all RF lines, except for the TX cancellation line, which was LMR-240. The radar system is powered by a triangular solar array with 1.2\,kW nominal power per face, and a 200\,Ah, 24\,V battery buffer. GPS provides accurate timestamping of events, and a WLAN bridge provides telemetry and commanding from Summit Station. 

The FPGA firmware included a software interface for debugging and diagnostic information as well as commanding. The interface allows us to: change the TX frequency and modulation, steer the polar angle of the phased transmitter, send software triggers, run the TX cancellation routine, change the attenuation setting on the DAC channels, enable/disable the RF trigger and set the threshold, select the number of surface stations required to form a trigger, set the delay time between a trigger and an event readout (to ensure enough data has been buffered before readout) and request different levels of debug and diagnostic data. Changing the TX frequency and modulation allows us to observe the reflected signal in different bands. The carrier cancellation finds the optimal amplitude and timing offset to cancel the transmitter in each receive channel simultaneously.  A single event record is a header with useful housekeeping information and 4 records of 32,768 bytes, corresponding to 16,384 samples of data, one per RX channel. Data is readout to a raspberry pi single board computer (SBC) via ethernet for fast transfer and recorded redundantly on two disks before telemetry over WLAN to Summit Station, where it is redundantly recorded once more and transmitted to servers in the US. A power distribution system allows for on/off control of the individual power amplifiers via software.

\subsection{Surface System} A single surface station consists of 2 scintillator panels\cite{IceCube:2012nn}, a cross-pol log periodic dipole antenna\cite{SKA:2018ckk}, a power system consisting of an omnidirectional 4x20\,W solar array with a 10\,Ah, 12\,V battery buffer, a raspberry pi SBC with redundant storage, and electronics for readout of the scintillators and radio antenna. The SBC controls the readout electronics for the scintillators and radio antenna, the latter of which is repurposed from the CODALEMA experiment~\cite{codalema_extasis}. There are 3 such stations on a 40\,m radius from the TX. Communication with the surface stations is achieved via ethernet on a cat-7 cable, with trigger signals sent along a separate cat-7 cable. The scintillator panels were deployed on the surface, and the radio antenna was deployed atop bamboo poles to elevate it approximately 1\,m off the surface. Every cosmic ray, RF, and forced/software trigger is sent to every surface station, triggering an event readout of the surface radio. 

\begin{figure*}[ht]
  \centering
  \includegraphics[width=0.9\textwidth]{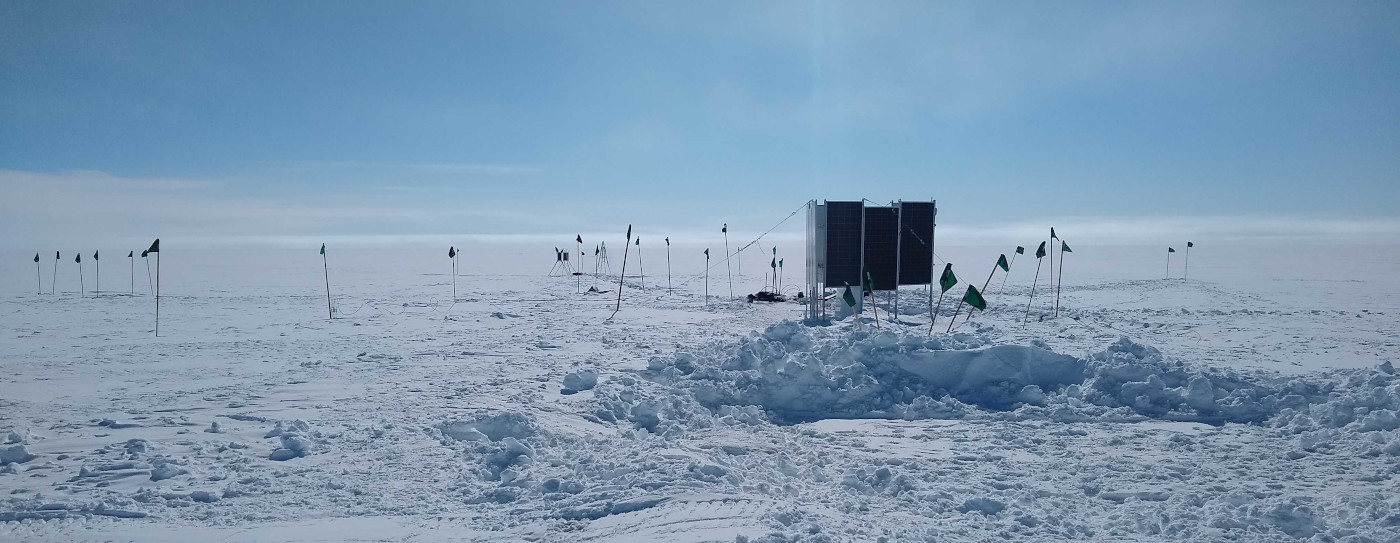}
    \caption{RET-CR, as deployed for the 2023 run. Solar panels are evident in the center, and flags denote the buried receivers and transmitter. In the middle distance are smaller solar trees and a surface radio antenna.}
  \label{deployed}
\end{figure*}

\section{Deployment} The deployment of RET-CR took place in May 2023 with a team of five people. RET-CR is located approximately 6\,km N/NE of Summit Station, Greenland. Summit Station is ideally suited to RET-CR, situated at 3,216\,m above sea level at roughly the center of the Greenland Ice Sheet (72.579583, -38.459186). This high elevation, uniform ice facilitates the deposition of a significant fraction of the primary cosmic ray energy into the ice, ideal for maximizing chances of measuring a radar echo. 

Station staff assisted in setting up a shelter/work tent at the site. The four boreholes were drilled using a Kovacs Mark-V coring system. The surface stations, including antennas, were assembled and deployed. The radar power system was assembled on site and raised, using guylines set to dead-men anchors to protect from the wind. The WLAN link was established by facing one point of the PV triangle directly toward Summit Station and mounting the WLAN antenna at this point. The DAQ, a single Nanuk 975 enclosure with custom bulkhead panels and inlet/outlet vents for thermal regulation, was buried in a 2 x 2 x 1\,m snow pit vault covered with plywood, to protect it from blowing snow. A photograph of the deployed instrument is shown in Fig.~\ref{deployed}, with the PV array visible at center right, and one of the surface stations in the middle distance. Flags indicate buried elements such as the receiver strings (far right) and cables. The large snow disturbance in the foreground was drifting against our deployment tent, which Summit Station staff had removed at the end of our deployment, just before this photo was taken.

\section{Initial performance} The instrument worked nominally for several hours as soon as it was powered on. By the next day, we noticed that two of the surface stations had powered off. It was concluded that the surface radio electronics were perhaps drawing too much power, and so these were unplugged from the systems for the duration of the deployment. During the commissioning phase one of the surface station trigger lines went offline, though we still had contact with that station via ethernet, and could verify that it was still triggering nominally. Therefore, all of our data was with N=2 surface station logic, though the timestamped triggers for the 3rd station were also recorded and stored. Additionally, only 4 transmitter channels were powered on at any time, as we noticed very quickly that the DAQ became far hotter than expected with all 8 transmitter power amplifiers active. 

After a 4 week commissioning period, during which the various subsystems were tested and firmware upgrades were written and pushed, a data run commenced. Communication with RET-CR was lost 9 days into this data taking run. Subsequent debugging efforts concluded that the system was still powered on, but that the WLAN link was down, and the system was not fully operational, as evidenced by an inability to detect the TX signal from the surface. The DAQ vault was then uncovered to reveal that overheating resulted in the DAQ box melting down approximately one meter into the snow, and severing the WLAN cable, as well as one of the surface station trigger lines, which had become unresponsive some days before. The DAQ itself had reset into a safe mode during the 9th day of the data run and was undamaged. 

\begin{figure}[h]
  \centering
  \includegraphics[width=0.5\textwidth]{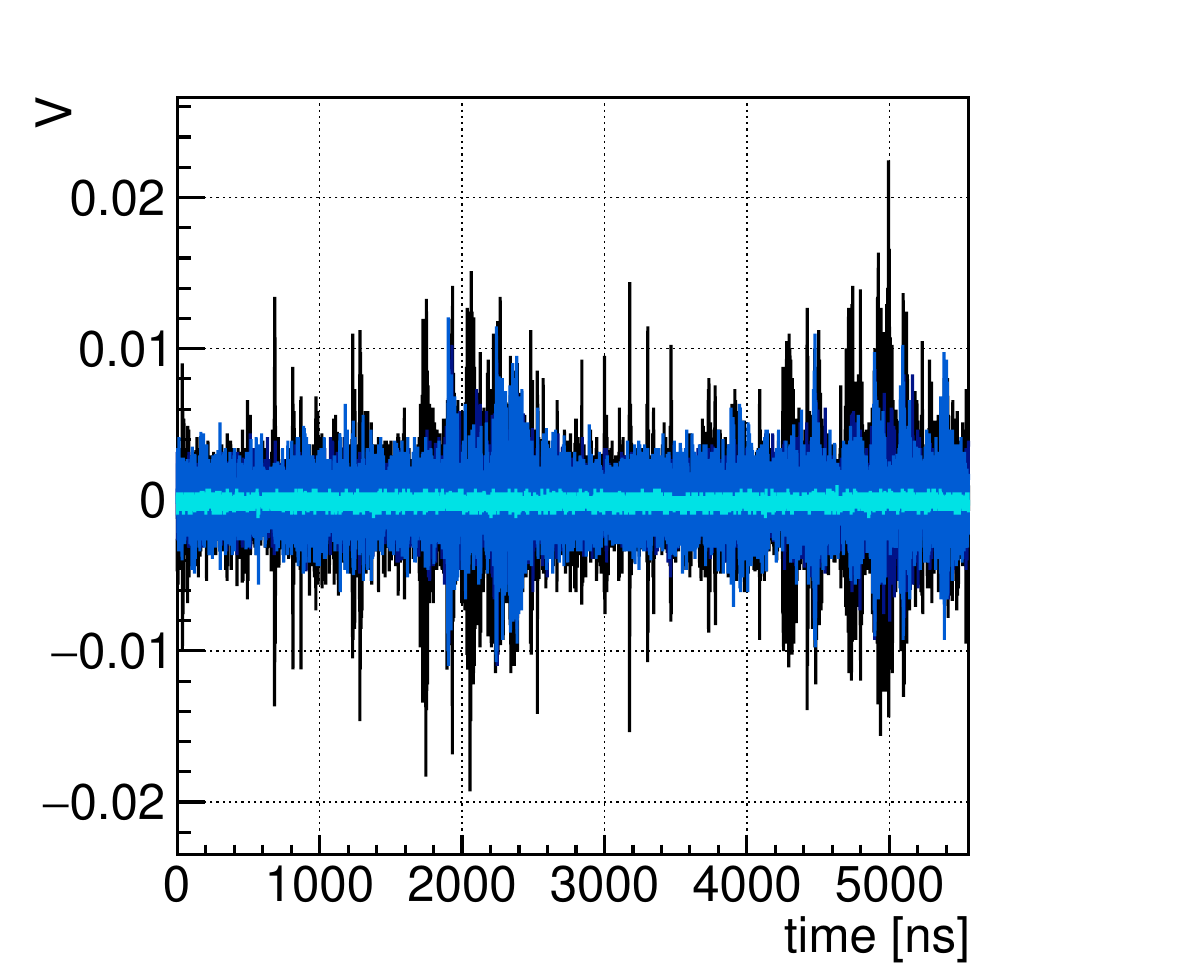}
    \caption{An example forced trigger event. Data from all 4 receiver channels are overlaid, the smallest trace being that of an unamplified surface antenna, the other three being the dedicated radar receiver channels. Evident in the data are broadband RF noise spikes from inadequately shielded RF power supplies, which were remedied for the 2024 data run.}
  \label{data}
\end{figure}

\begin{figure}[h]
  \centering
  \includegraphics[width=0.5\textwidth]{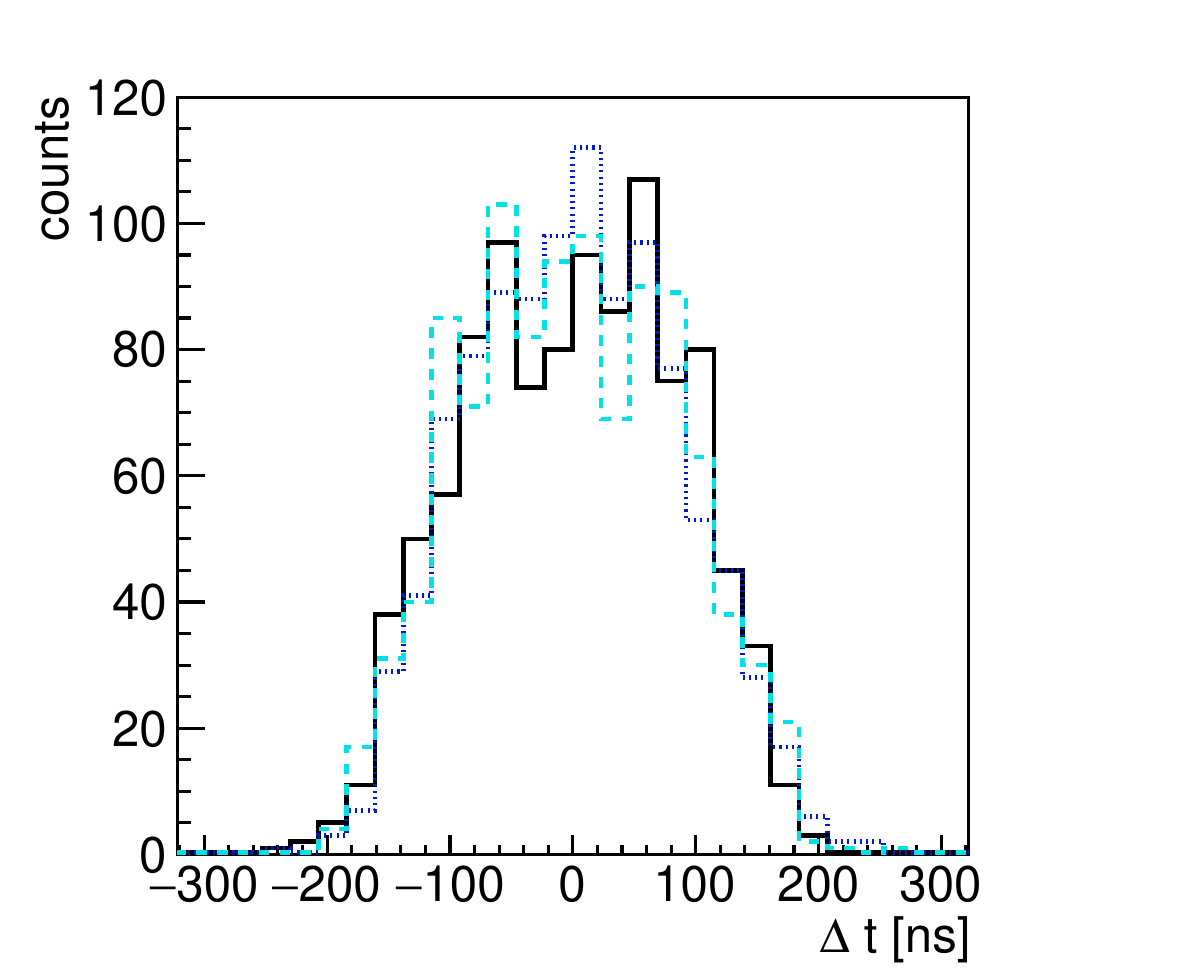}
    \caption{The time difference in trigger arrival times between pairs of surface systems (each line style/color represents a surface system pair), as recorded on the central radar DAQ. Direct line separation of approximately 70\,m sets the maximum light crossing time.}
  \label{dt}
\end{figure}

Our 220 hours of livetime demonstrated several important technical successes. The first and most important of these was the transmitter cancellation. Without active cancellation, the receiver amplifiers would be fully saturated by the direct transmitter signal. To remedy this, a time delayed and scaled version of the transmitted signal is fed directly into the RX channels before the front-end amplifier to cancel out the incoming TX signal. A firmware/software routine periodically runs to find the optimal cancellation phase and amplitude. The phase and amplitude that are selected are those which optimally cancel out the received transmitter signal. This is the sum total of the direct TX signal, plus all reflected paths from internal reflecting layers, and most prominently, the surface. Therefore, the cancellation phase will vary as the surface profile changes through wind or precipitation. Fig.~\ref{windspeed} shows this qualitatively. Here the magnitude of the change in cancellation phase $|\Delta \phi|$ between hourly runs of the TX cancellation in one receiver is plotted against the local windspeed, as measured at Summit Station. Higher windspeed results in increased drifting, which changes the reflected path lengths for the TX signal, thus requiring a more significant change in cancellation phase. 

Even at the relatively modest power of 20\,W per channel, the signal received directly by the RX was far higher than the dynamic range input of the front end amplifiers. We were able to use the TC to nearly completely eliminate the TX signal from the RX, even with the TX at full power. The TC happens in two stages. First, the transmitter is set to fractional power, so that the RX amplifiers do not saturate, and the attenuators on the TC DACs are set to a high value. A first stage cancellation is run, to find an approximate value for the cancellation phase and amplitude. Second, the transmitter is increased to full power and the attenuation on the TC DACs are reduced appropriately to best match the amplitude of the incoming signal. Then the second stage TC routine is run, to fully eliminate the signal. Consequently, a plot comparing TC on versus off cannot be shown, as turning the TC off would damage the receivers. We can approximate the cancellation by a simple calculation: a 20\,W output, assuming zero gain to the TX and RX, results in a direct TX signal of $\sim$100\,mV at the RX. When the TC is run, the TX is removed to the level of noise, which is nominally $\sim$10\,$\mu$V thermal for our band, equating to roughly 80\,dB of rejection in power.

Steering of the beam was demonstrated to work as expected by varying the relative phase of the transmitter antennas and observing the changing amplitude in the receiver strings. Fig.~\ref{dt} shows the arrival time difference between pairs of surface system triggers for a period during commissioning when N=3. The straight-line distance between stations is roughly 70\,m, corresponding to a maximum light crossing time of approximately 230\,ns. After debugging during commissioning, the data readout was also nominal for the data run, with readout taking $\sim$50\,ms per event. 

Another issue that became evident during the run was regular broadband radio frequency interference (RFI) that contaminated the data, visible in Fig.~\ref{data}. We subsequently identified the source of this RFI to be the high current switching power supplies feeding the transmitter amplifiers, which were removed entirely for the subsequent 2024 run. 

\section{Improvements for 2024} The most critical improvement made after the 2023 run was a redesign of the thermal regulation system. Our design of this initial system was predicated upon a surface deployment, and it allowed us to shed heat sufficiently in such a scenario. However, on arrival at Summit we realized that the filtration system we had designed to keep snow out of the inlet/outlet ports would be insufficient for the local conditions: the inlet filter was ineffective against the finest snow, which could damage/destroy the electronics, and the outlet filter could clog/freeze without regular maintenance. We therefore elected to bury the system, which inhibited cooling, and ultimately caused the system to melt down into the snow. To mitigate this, the 2024 enclosure has an improved, passive thermal regulation system to meet the local conditions. The power amplifiers are bolted to the inside of a large heat sink that forms the lid of the revised electronics enclosure, which was deployed on a platform above the snow surface for 2024.

Other improvements include reconnecting the surface system radio receivers for improved reconstructability of the primary cosmic rays. To achieve this, we doubled the capacity of the surface system power supply. The 2024 system also has two additional surface stations, for a total of 5, with the two additional stations at longer baselines. As deployed in 2023, 3 RX channels had the full front end chain, with a 4th channel connected, but not buried, and without any front end electronics. The 2024 system included an additional amplified, buried channel, opting to break the symmetry of the layout and deploy one RX at a longer baseline. Finally, the 2024 system has different power regulation to completely remove the main sources of local RFI.  

The 2024 data season from May-August has just concluded. The improvements allowed for the system to operate all through the summer, capturing $\mathcal{O}(10^5)$ cosmic ray triggered events. Each event record in the 2024 system contains the data from 4 amplified channels (three with active carrier cancellation), one unamplified monitoring channel, and 5 surface stations including the data from surface radio antennas and two scintillator panels per station. Analysis of these data, and details of the improved hardware and firmware, will be presented in forthcoming publications. 

\section{Conclusions} We have presented the first deployment of RET-CR, and the initial performance of the instrument. RET-CR was deployed for the first time in May 2023, and took 9 days of data after a four week commissioning period. The instrument performed well overall during the data taking period, before going offline, though the data was contaminated by broadband RFI. All problems that were identified during the run were remedied in advance of the data run from May-August 2024, which has just concluded. 
\\
\\
\section{Acknowledgments} We sincerely thank the support staff at Summit Station for their tremendous efforts in getting RET-CR into the ice safely and successfully. We also thank K.~Hughes and the RNO-G collaboration for their assistance. We thank M.~Kauer, C.~Wendt, D.~Tosi, and the IceCube collaboration for providing our scintillator panels and associated technical support. We thank the CODALEMA experiment for providing our surface radio electronics. We recognize support from The National Science Foundation under grant numbers 2012980, 2012989, 2306424, and 2019597 and the Office of Polar Programs, the Flemish Foundation for Scientific Research FWO-G085820N, the European Research Council under the European Unions Horizon 2020 research and innovation programme (grant agreement No 805486), the Belgian Funds for Scientific Research (FRS-FNRS), IOP, and the John D. and Catherine T. MacArthur Foundation.

\bibliography{bib}
\end{document}